# Model Reuse through Hardware Design Patterns

Fernando Rincón    Francisco Moya    Jesús Barba
Juan Carlos López
University of Castilla-La Mancha
13071 Ciudad Real, Spain
{Fernando.Rincon, Francisco.Moya, Jesus.Barba, JuanCarlos.Lopez}@uclm.es

## Abstract

*Increasing reuse opportunities is a well-known problem for software designers as well as for hardware designers. Nonetheless, current software and hardware engineering practices have embraced different approaches to this problem. Software designs are usually modelled after a set of proven solutions to recurrent problems called design patterns. This approach differs from the component-based reuse usually found in hardware designs: design patterns do not specify unnecessary implementation details.*

*Several authors have already proposed translating structural design patterns concepts to hardware design. In this paper we extend the discussion to behavioural design patterns. Specifically, we describe how the hardware version of the Iterator can be used to enhance model reuse.*

## 1. Introduction

A significant increase in design productivity is usually a direct consequence of two key factors: 1) raising the abstraction level, and 2) reusing previous designs.

Design reuse is usually achieved using IP-based methodologies (hard, firm and soft IP [1]). IP blocks offer closed solutions and limited flexibility (parameterization or configuration). They are usually optimized under some assumptions and modifications are either impossible or extremely tricky, with the exception of some extensible IPs such as configurable processors which are designed with user modification in mind. Besides, in most cases these IP blocks are completely designed at the RT level using an HDL.

Most often system designers would like to reuse behavioural level abstractions (algorithms) while IPs usually just offer implementations of such algorithms under a certain set of constraints. The solutions proposed to this problem range from behavioural level modeling to high level synthesis. Unfortunately, these higher level tools may introduce large overheads in terms of resource allocation and performance of the final solution.

In this paper we propose a focus shift from component-level reuse as advocated by IP-based methodologies to model-level reuse as represented by current software engineering practice. Our point of view improves reusability at two levels: 1) at the system-level, using pattern oriented modeling. Most design patterns are not related to a particular implementation. 2) At the component-level, introducing a broader sense of genericity than the only currently available in HDLs, in order to allow generic algorithm descriptions.

There has been a lot of discussion on the technical problems of component-level reuse, the key goal of most IP-based methodologies. Some of these problems are still open to discussion: 1) integration of components in the design flow, 2) interface compatibility and generation, and 3) customization and optimization. The IP designer is faced with these issues very early in the system design cycle, with the risk of precluding reusability under unforeseen scenarios.

Published literature proposes partial solutions for each of the above mentioned issues: 1) IP-based methodologies ease the integration of third party IPs into the design flow, 2) interface compatibility is usually handled through standardization [1] and wrapper generation [6], and 3) behavioural IPs [8] and high-level synthesis provide the ability to adapt IPs to new scenarios.

Nevertheless, reusing a predefined component may be the source of area and performance overheads due to either interface compatibilization or unnecessary IP features. The latter is not uncommon since IP reusability is usually improved by considering several target scenarios. Modifications of third party IP are either impossible or extremely tricky. In this paper we propose model reuse in contrast to component reuse.

Most of current IPs are designed at the RT-level using a conventional hardware description language. Raising the level of abstraction would provide two important benefits: 1) models would be easier to modify and maintain, and 2) it would be possible to reuse the primitive modeling elements

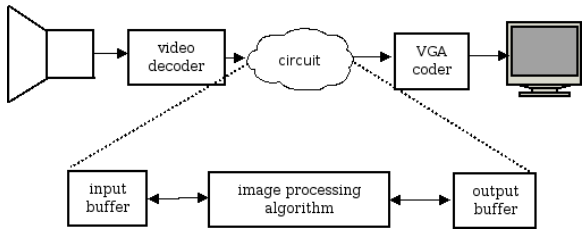

**Figure 1. Design example**

(algorithms, data structures, etc.).

In the following sections, first we will present the motivating example showing the limitations of current design methodologies. Next we will introduce the problem of model reuse versus component reuse. We will then outline the solution we propose which is based on a hardware version of an existing behavioural software pattern: *the Iterator*. This will lead to the definition of a basic component library that will later be used to reengineer the motivating example. In section 3.4 we will describe implementation details of this pattern. Some experimental results will be shown in section 4. Finally, we will present some conclusions and draw future related work.

## 2. Embracing change

Let us consider the design of a simple real-time image processing application. The whole system would be composed of the entities shown in figure 1: camera, video decoder, image processing circuit, VGA coder, and video monitor.

The image processing circuit would be composed of an input buffer to acquire the video stream, an image processing algorithm and an output buffer to accommodate the output video stream. For the sake of clarity, we will first consider the case of the simplest possible algorithm: copying data from the input buffer to the output buffer. Such a simple algorithm may be implemented as a finite state machine handling the buffer signals and sequencing the read and write operations.

Let's suppose there is a small change in the system. The input video stream is now fed into a RAM, storing a whole frame instead of using a sequential access buffer. The stream copy algorithm does not change at all: we still need to transfer data from the video source to the video sink. But the implementation must be radically changed to accomodate the new interface requirements of the memory. Now there is a need to maintain a memory address register pointing to the appropriate position in RAM. A similar situation arises if we change the output buffer to the VGA decoder into another RAM block.

Pixel format is another kind of critical modification when trying to reuse a system like the one described. Changing 8-bit grayscale format into a 24-bit RGB format, for example, may require non-trivial modifications in the state machine specification, since obtaining each pixel may require more than one memory access.

Although it seems feasible to reuse the copy algorithm in many scenarios, it is not so easy because of the coupling between algorithms, data structures and hardware interface handling.

In this paper we propose to decouple algorithms and data structures by means of a hardware version of a well known design pattern: the iterator pattern [7]. The main goal of this pattern is to avoid exposition of the internal implementation when accessing a data structure. The hardware interface of data structures will only be exposed to iterators and components which implement those data structures. Both are either automatically generated or reused. The impact on the overall design of modifying hardware interfaces is thus minimized.

## 3. Model Reuse

Increasing reuse opportunities is a well-known problem for software designers as it is for hardware designers. Nonetheless current software and hardware engineering practice embraced different approaches to this problem. Software designs are usually modeled after a set of proven solutions to recurrent problems called design patterns. This approach differs from the component-based reuse usually found in hardware designs in that the design patterns do not specify unnecessary implementation details. Therefore, design patterns allow the reuse of higher level abstractions in a wide variety of scenarios without the performance overhead of behavioural IPs.

Several authors have already proposed translating some of the design patterns concepts to hardware design [6, 12, 4] but to our knowledge all previously published works are entirely devoted to structural and creational patterns, already similar to current hardware design practice. In this paper we introduce behavioural patterns in the domain of hardware design, allowing a more abstract view of the design.

We do not advocate a complete parallelism between software and hardware design process. Many of the most successful design patterns do not have a hardware counterpart. Therefore there is a need to develop a hardware version of a design pattern catalog, similar to what is already available in software [12]. Similarly, there is a strong need to develop standardized foundation libraries combining the most successful patterns in an integrated way.

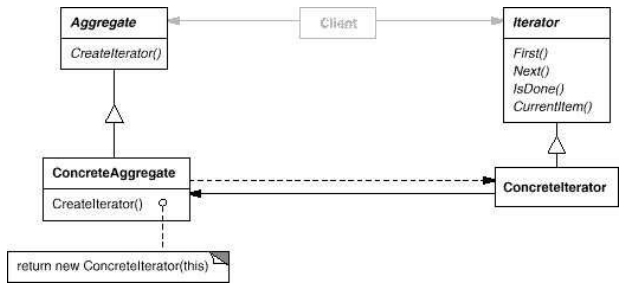

**Figure 2. Iterator pattern**

## 3.1 The iterator pattern

The *iterator* [7] design pattern (figure 2) is already found in most software foundation libraries. Iterators provide a way to access the elements of a data store (aggregate object) without exposing its undelying representation.

The hardware version of entity *Iterator* defines an interface for accessing and traversing elements, as shown in table 2. *ConcreteIterator* implements the *Iterator* interface and keeps track of the current position in the traversal of the *Aggregate*. Due to the static nature of hardware some of the details of the software *Iterator* are not applicable. For example, the Aggregate is not responsible for creating *Iterator* objects. Iterators must be instantiated at design time.

## 3.2 Basic Component Library

In a hardware design, data storage is based on a wide range of memory technologies. Some require sequential access, others are accessed randomly, or even associatively. Besides, there are lots of algorithms performing common data manipulation (copy, transform ...). Others are specific to an application domain (e.g. pixel-wise filtering, convolution filtering or binary image labelling for image processing applications). All those elements should be packed into a library of basic components, in a similar way to software foundation libraries. Designs built over this kind of libraries are proven to easier to develop and maintain, and less error-prone.

In the remaining of the current section, we will introduce a proposal of such basic library, inspired on the C++ Standard Template Library [11]. This library is organised around three kinds of concepts: containers, iterators and algorithms.

### 3.2.1 Containers

Containers are equivalent to the Aggregate objects in the Iterator Pattern. They are collections of elements that can be implemented in a number of physical structures. Containers are accessed through iterators. Table 1 classifies the basic set of containers depending on the type of memory access required (random or sequential), and the type of traversal allowed (forward, backwards or both).

**Table 1. Common containers**

| Containers | Random | | Sequential | |
|---|---|---|---|---|
| | Input | Output | Input | Output |
| stack | - | - | F | B |
| queue | - | - | F | F |
| read buffer | - | - | F | - |
| write buffer | - | - | - | F |
| vector | ✔ | ✔ | F, B | F, B |
| assoc. array | ✔ | ✔ | - | - |

**Table 2. Iterator Operations**

| Operation | Meaning | Applicability |
|---|---|---|
| inc | move forward | F / F, B |
| dec | move backwards | B / F, B |
| read | get the element | random / F, B |
| write | put the element | random / F, B |
| index | set the current position | random |

### 3.2.2 Iterators

Table 2 shows the set of operations allowed for each type of iterator (forward, backwards or bidirectional). All iterators keep track of their current position in the traversal of the container. They are also able to read and/or write the element at that position. Forward iterators include an additional operation to advance the current position. Backward iterators include the possibility to move backwards. Random iterators can set arbitrary positions through the index operation. Although the iterator provides a common interface for any container, it must have a deep knowledge of the internals of the container. For that reason a concrete iterator must exist for each type of container in the library.

### 3.2.3 Algorithms

The basic components library should also include a set of commonly used algoritms. Every one should use the interface provided by iterators to access data in the containers. This would guarantee reusability of the algorithm, despite of the container chosen for a certain implementation. It is out of the scope of this paper to delimit the set of algorithms that should be included in the basic library. For illustration purposes we have just selected two simple ones: copying data from a source to a destination, and an image blur filter.

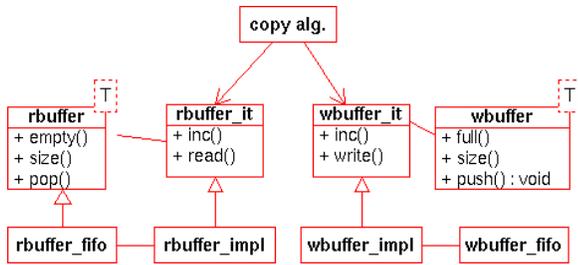

**Figure 3. Model of the example**

### 3.3 The Example Revisited

Lets apply the iterator pattern to the example described in section 2. Figure 3 shows the resulting model. Now data acquisition from the video decoder has been modelled as a read buffer container (*rbuffer*), while the output video stream is fed into a write buffer container (*wbuffer*). Access to *rbuffer* and *wbuffer* containers is abstracted through *rbuffer_it* and *wbuffer_it* iterators respectively. Therefore, the video processing algorithm avoids direct manipulation of the containers. As seen in figure 3, containers are implemented over FIFO structures (note *rbuffer* and *wbuffer_fifo* suffixes) that accomodate the data stream. The copy algorithm is almost trivial: an endless loop that sequences read and write operations and iterator forwarding for both containers. All these operations can be performed in parallel in a hardware implementation. Note that the simplicity of the algorithm does not affect the validity of our approach, that can be generalized to any other video processing algorithm.

Lest's suppose that the system must be modified for a new configuration, where both input and output streams are fed into two separate static RAMs. This change does not really affect the model, since the specific aggregates remain the same (wbuffer and rbuffer). Therefore, the other elements in the model are not affected either, and no modifications are needed. The only difference is that the aggregates must be implemented over static RAM instead of on-chip fifos. This implementation can be automatically generated as described in section 3.4, so this kind of modifications requires no design effort.

It would also be possible to modify the pixel data representation (from 8-bit grayscale to 24-bit RGB, for example). Here two different alternatives arise depending on the RAM data bus size: 1) For a 24-bit data bus, we should only re-generate the implementations of the elements using the 24-bit data pixel as the base type. 2) For an 8-bit data bus, we should also modify the iterator code to perform three consecutive container reads/writes to get/set the whole pixel. In any case, all this scenarios can be considered by the automatic code generator, thus requiring no designer intervention.

### 3.4. Pattern Implementation

Most design patterns are naturally described using the object-oriented (OO) paradigm . While there exist several hardware description languages with a limited support of OO concepts (e.g. SUAVE [3], SystemC [2]), synthesis tools that are able to generate a final implementation from a hardware OO model are not mature at all, despite this is an active research area [9].

On the other side, traditional hardware description languages, such as Verilog or VHDL, do not support some of the necessary abstractions. Direct translation of the proposed pattern would require at least: inheritance support, a broader sense of genericity, and support for partial template specialization.

Our solution is based on the concept of metaprogramming [10]. An automatic code generator produces customized versions of containers and iterators from a code template. The template includes information on the available operations, shared resources and parameterized code fragments. The result is a set of efficient VHDL components, ready to be synthesized. Algorithms can be also described through metamodels, although they have not been considered in this paper and are left as future work. The code generator must also be able to implement communication mechanisms for complex data types whose sizes do not match that of the ports between components (such as the modification of the pixel format discussed in 3.3). It is also the responsible for including only those resources that are really used by the selected operations.

Metaprogramming provides a number of additional benefits. It allows automatic generation of arbitration logic for shared physical resources (e.g. RAM). It also provides transparent selection of the communication protocol between components. Here transparecy refers to the model, not to the designer that must select the right values for the different parameters considered in the metamodel.

As outlined in section 3.2, containers may be mapped to several physical devices. All of them can be implemented in any kind of RAM memory, while stacks can also be implemented over FIFO cores, or queues and read/write buffers can also mapped over LIFOs (these cores are commonly found in FPGA designs).Some targets provide the most efficient implementation (such as a queue over a FIFO core), while others may lower the overall system cost (such as the same queue over an external RAM).

Metaprogramming defers until the last moment the selection of the proper implementation of a container, depending on the requirements of the application. However, this

```
entity rbuffer_fifo is
  port (
    -- methods
    m_empty : in  std_logic;
    m_size  : in  std_logic;
    m_pop   : in  std_logic;
    -- params
    data    : out std_logic_vector(7 downto 0);
    done    : out std_logic;

    -- implementation interface
    p_empty : in  std_logic;
    p_read  : out std_logic;
    p_data  : in  std_logic_vector(7 downto 0)
    );
end rbuffer_fifo;
```

**Figure 4. Read buffer over a FIFO device**

```
    ...
    -- physical interface
    p_addr : out std_logic_vector(15 downto 0);

    p_data : in  std_logic_vector(7 downto 0);

    req : out std_logic;
    ack : in  std_logic
    );
end rbuffer_sram;
```

**Figure 5. Read buffer over an SRAM device**

selection may be based on previous characterization of the design space. For example, in this paper, we characterized all the physical devices available in the target platform (the XSB-300E prototype board from XESS ). We obtained information about data access times for every container, area, power comsumption . . . . Since components are generated automatically, it is feasible to generate versions of each one for every physical target and range of configuration parameters. This characterization of the design space would delimit the region of interest given a certain set of constraints.

Only specific classes have a physical counterpart, when mapping the abstract model of the design pattern over the implementation. Each one becomes a VHDL entity and an architecture. Abstract classes do only exist inside the domain of the model, and are used to define a common functional interface. The physical interface (that of the VHDL entity), includes not only the functional interface, but also an extra implementation interface, that includes, among others, all the ports to interface the physical devices. The physical entity of a container inmplemented over a static RAM, for example, will include a port for each operation and each parameter from the functional interface (read, empty . . . ), in addition to all the ports related to the SRAM interface (p_addr, p_data . . . ). The VHDL architecture of the container will define the logic to perform the mapping.

Figures 4 and 5 show the entities implementing the container *rbuffer* used in the example in section 3.3. The first one belongs to the implementation over a FIFO, while the second one includes only the differences (the implementation interface) with respect to the first.

The VHDL architecture is simply a wrapper of the FIFO core in the first case, and hardly includes any logic. In the second case, the architecture encloses a little finite state machine that controls memory access, as well as a few registers to store the begin and end pointers of the queue (implemented as a circular buffer) over the static RAM.

Iterators are also generated from their own metamodel. One iterator metamodel must be defined for each kind of container. The metamodel is the responsible for providing the necessary functionality, depending on the type of traversal and on whether it is an input or an ouput iterator. The iterators used in the previous example don't include much functionality since they are extremely simple. In fact they are no more than a wrapper that renames some signals and provides the common interface already mentioned.

The implementation of the copy algorithm is really simple, since it only has to activate the *inc* and *read* operations for the input iterator, and the *inc* and *write* operations for the output iterator. The data is also connected from the input to the output.

## 4. Experimental Results

The numbers in table 3 show the results of the implementation of three design examples using the Iterator pattern. As a target platform we use the XSB-300E board from XESS. Each cell shows the results corresponding to the pattern-based versus the custom implementation. The first two rows correspond to the example revisited in section 3.3. The third example is a little bit different. In this case, we have implemented a blur filter that processes an image coming from the video decoder and sends it to a VGA coder. The *rbuffer* container, instead of a simple FIFO has been mapped over a special one. It is a 3-line buffer structured to provide 3 pixels in a column for each access. This makes the convolution product in the blur algorithm very simple and quite efficient since ideally a new filtered pixel can be generated at each clock cycle.

It is clear from the results in the table, that there is a negligible overhead for the pattern-based implementation. The use of the pattern affects only the structure of the model, but does not necessarily imply extra logic. This is for example the case of the iterators, which are only wrappers that will be dissolved at the time of synthesizing the design. In the above examples, ad-hoc solutions contain almost the same physical components (FIFOs, algorithm FSM . . . ).

The saa2vga examples represent two different points of the design space. The first one (the FIFO implementation) provides maximum performance at the highest cost. The SRAM implementation is much smaller, but performance will depend on memory access times.

**Table 3. Design experiments.**

| Design | FFs | LUTs | block RAM | clk MHz |
|---|---|---|---|---|
| saa2vga 1 | 147/147 | 169/168 | 2/2 | 98/98 |
| saa2vga 2 | 69/69 | 127/127 | 0/0 | 96/96 |
| blur | 3145/3145 | 4170/4169 | 2/2 | 98/98 |

## 5. Conclusions and Future Work

The use of IPs has proven to be a very effective way of managing the ever growing complexity of SoCs. However, system integration is only a part of the design problem. Higher level design methodologies, where solutions are considered in the domain of the problem without being conditioned by the implementation must be provided. This is recently attracting the interest of the research community to some of the most successful paradigms used in software design, such as OO Programming, UML and design patterns.

The application of such technologies is not straightforward, due to the important differences between both domains, but some contributions have been done to the use of UML for system level design[5, 13], or the synthesis from an object-oriented specification [9]. Also some papers have already considered the applicability of design patterns to hardware design. However, the fact that design patterns are based on the object-oriented paradigm, and the lack of a mature technology for OO-synthesis, makes that only structural patterns are considered, since they are easily mapped over a hardware description language.

In this paper we propose the use of a well known software design pattern (the Iterator) to provide a way to decouple algorithms from the underlying data structures. This would make algorithms more generic and thus provide more opportunities for later reuse. Unlike previous proposed hardware design patterns, the iterator is behavioural, and requires a higher level of abstractions that the one provided by HDLs.This have been solved through the use of metaprograming. Also a basic components library based in this pattern has been proposed. This library provides eficient physical implementations for sequential and random data structures. The deliberate simplicity of the examples shown in the paper proves the flexibility and low overhead provided by this modelling approach.

Along the paper, we raised a number of questions that must be addressed. There is a need to develop a hardware version of a design pattern catalog, similar to what is already available in software. Specific application domains such as video image processing demand specific libraries including common algorithms (convolution filters, image labelling ...). and specialized iterators. A hardware abstraction layer should be considered for the development of the metamodels.